\documentclass[usenatbib]{mn2e}
\usepackage[english]{babel}
\usepackage{amssymb}
\usepackage{multirow}
\usepackage{graphicx} 
\usepackage{float}
\usepackage{nameref}
\usepackage{footmisc}
\usepackage{wrapfig}
\usepackage{wasysym}
\usepackage{times}
\usepackage{textcomp}
\usepackage{epsfig}
\usepackage{natbib}
\title[CXOU\,J1647$-$45 Outburst Pulse Phase Analysis]{Pulse Phase-coherent Timing and Spectroscopy of CXOU\,J164710.2$-$45521 Outbursts}
\author[Rodr\'iguez Castillo G. A. et al.]{Guillermo A. Rodr\'iguez Castillo,$^{1,2}$ Gian Luca Israel,$^2$ Paolo Esposito,$^3$ Jos\'e A. Pons,$^4$
  \newauthor  Nanda Rea,$^{5,6}$ Roberto Turolla,$^{7,8}$ Daniele Vigan\`o,$^{4,5}$ and Silvia Zane$^8$\\
  $^1$Dipartimento di Fisica, Sapienza Universit\`a di Roma, Piazzale Aldo Moro 5, 00185, Rome, Italy\\
  $^2$INAF-Astronomical Observatory of Rome, via Frascati 33, 00040, Monte Porzio Catone, Italy\\
  $^3$INAF-IASF Milano, Via E. Bassini 15, I-20133 Milano, Italy\\
  $^4$Departament de F\'isica Aplicada, Universitat d'Alacant, Ap. Correus 99, 03080, Alacant, Spain\\
  $^5$Institute of Space Sciences (CSIC-IEEC), Campus UAB, Faculty of Science, Torre C5-parell, E-08193 Barcelona, Spain\\
  $^6$Astronomical Institute ``Anton Pannekoek", University of Amsterdam, Postbus 94249, 1090 GE Amsterdam, The Netherlands\\
  $^7$Department of Physics, University of Padova, Via Marzolo 8, 35131, Padova, Italy\\ 
  $^8$Mullard Space Science Laboratory, University College London, Holmbury St. Mary, Dorking, Surrey, RH5 6NT, UK}
\def\sgr {SGR 0418+5729}

\def\src{CXOU J164710.2$-$455216}

\def\swift{{\em Swift}}
\def\XMM{{\em XMM-Newton}}
\def\chandra{{\em Chandra}}
\def\ergscms{erg cm$^{-2}$ s$^{-1}$}
\def\ergs{\rm erg/s}
\def\cxou{CXOU J1647-45}

\begin{document}
\date{MNRAS, published online May 8, 2014. doi: 10.1093/mnras/stu603}
\maketitle
\begin{abstract}
We present a long-term phase-coherent timing analysis and pulse-phase 
resolved spectroscopy for the two outbursts observed from the transient 
anomalous X-ray pulsar \src. For the first outburst we used 
11 \emph{Chandra} and \XMM\ observations between September 
2006 to August 2009, the longest baseline yet for this source. 
We obtain a coherent timing solution with $P=10.61065583(4)$ s, 
$\dot{P} = 9.72(1) \times 10^{-13}\;$s s$^{-1}$ and $\ddot{P} = 
-1.05(5)\times10^{-20}\; $s s$^{-2}$. Under the standard assumptions 
this implies a surface dipolar magnetic field of $\sim 10^{14}$ G, 
confirming this source as a standard $B$ magnetar. 
We also study the evolution of the pulse profile (shape, intensity
and pulsed fraction) as a function of time and energy. 
Using the phase-coherent timing solution we perform a phase-resolved 
spectroscopy analysis, 
following the 
spectral evolution of pulse-phase features,
which hints at the physical processes taking place on the star. 
The results are discussed from the perspective of magneto-thermal 
evolution models and the untwisting magnetosphere model.
Finally, we present similar analysis for the second, less intense, 
2011 outburst.
For the timing analysis we used \emph{Swift} data together with 2 
\XMM\ and \emph{Chandra} pointings.
The results inferred for both outbursts are compared and briefly discussed 
in a more general framework.
\end{abstract}


\section{INTRODUCTION}
Soft $\gamma$-repeaters (SGRs) and anomalous X-ray pulsars (AXPs) 
are isolated neutron stars (INSs) with prominent high-energy
manifestations. They are characterized by 
rotational periods in the 
0.3--12\,s range and period derivatives (usually) larger than those 
typical of the radio-pulsar population ($\dot{P}\sim10^{-13}-10^{-10}$s/s). 
They exhibit peculiar flaring activity (see e.g. Mereghetti 2013) over a 
large range of time-scales (milliseconds to minutes) and luminosities 
($L\sim10^{38-47}$ erg s$^{-1}$).
Estimates of their magnetic field, derived under the usual assumptions for
isolated rotation-powered pulsars, place them at
the high end of the pulsar population ($B\approx 10^{14-15}$\,G). 
This, and other direct (Tiengo et al. 2013) and 
indirect evidences, suggests that they host an ultramagnetized 
neutron star, or magnetar (Duncan \& Thompson 1992, Thompson \& Duncan 1995).

Since the detection of SGRs/AXPs as persistent X-ray sources, one of the 
main concern has been the imbalance between the emitted luminosity
and the rotational energy loss rate, $\dot E$. Rotation is believed 
to be the standard mechanism that provides the energy output in canonical 
radio-pulsars. However, in SGRs/AXPs $\dot E$ is orders of magnitude 
below $L_{\mathrm{X}}$, although in some transient sources the 
rotational energy loss rate may exceed luminosity in the 
quiescent state (see e.g. Rea et al. 2012a).
Energy might be supplied by accretion, if a feeding companion is present
as is the case of many X-ray (binary) pulsars. Despite intensive searches, 
however, no binary companions have been detected so far around SGRs/AXPs
(see e.g. Woods et al. 2000; Camilo et al. 2006 for the tightest
constraints).

A more likely  alternative is that SGRs/AXPs are magnetically powered 
sources in which the decay/rearrangement of their (huge) magnetic field is 
responsible for both their persistent and bursting emission. Nowadays 
the magnetar model appears to be the more viable one and it will be assumed 
in this investigation, in particular for what concerns the timing and 
spectral analysis. In the following we shall 
refer to SGRs/AXPs as the ``magnetar candidates'', or simply as 
magnetars. Alternative scenarios have been proposed with varying degree of 
success to explain the SGRs/AXPs phenomenology, and include fallback discs 
(see e.g. Alpar et al. 2012), Thorne-{\.Z}ytkow objects (van Paradijs, Taam, 
\& van den Heuvel 1995), strange/quark/hybrid stars (see e.g. Horvath 2005; 
Xu 2007; Ouyed, Leahy \& Niebergal 2010) and fast rotating, highly-magnetized 
(B $\sim 10^{8-9}$ G) massive white dwarfs (Paczy\'nski 1990;  Malheiro, Rueda 
\& Ruffini 2012), among others (see Turolla \& Esposito 2013, section 5 and Mereghetti 
2008, section 7 for overviews).

\subsection{OUTBURSTS IN MAGNETARS}

Most of the known magnetar candidates are transient sources.  
A transient episode in a magnetar can be defined as an outburst
characterized by a rapid (minutes--hours) increase of the persistent 
flux by a factor of $\sim 10$--1000, with a subsequent decay 
back to the quiescent level on time-scales of months--years. Short 
bursts, which usually trigger detection, are emitted in the early 
phases of the outburst. Recurrent outbursts have been observed in 
a few sources (see Rea \& Esposito, 2011 for a review). 

Within the magnetar picture, outbursts occur quite 
naturally. According to our current understanding, one of the major 
differences between the magnetar candidates and pulsars is not (or not 
only) the higher value of the dipole field (there are low-field 
magnetars with $B\la 10^{13}$ G and high-field pulsars with $B\ga 10^{13}$ 
G), but the presence of a strong toroidal component in the internal 
field (Turolla et al. 2011 and references therein). 
It is the dissipation of the internal field which powers 
the magnetar bursting/oubursting behaviour by injecting energy deep in the 
star crust and/or by inducing displacements of the surface layers, with 
the consequent ``twisting'' of the external field (e.g. Thompson, Lyutikov 
\& Kulkarni 2002; Perna \& Pons 2011; Pons \& Rea  2012). The rate at 
which these episodes occur is different in different sources and is 
believed to depend mostly on the star magnetic field at birth and on its age.

Since the discovery the first confirmed transient magnetar in 2003 (XTE
J1810-197, which exhibited a flux enhancement by a factor of $> 100$; 
Ibrahim et al. 2004), outbursts have been the object of much interest. 
This stems from the possibility of testing, during the outburst decay, 
theoretical predictions over a relatively large luminosity
range in a single source, where a large number of important 
parameters are not changing, like, e.g., distance, mass, radius, age, 
viewing geometry (see e.g. Bernardini et al. 2009;
Albano et al. 2010; Rea et al 2013). 
\src\ is among the transients with the larger
flux variation. Following the outburst of 2006 September, in fact, its 
flux grew by a factor of $\gtrsim 300$ (Campana \& Israel 2006).

\begin{table*}
\renewcommand{\arraystretch}{1.3}
\resizebox{8cm}{!} {
\begin{minipage}{7.5cm}

\begin{tabular}{|l|l|c|c|c|}
\hline
\renewcommand{\arraystretch}{1.0}
\multirow{2}{*}{Telescope} & \multirow{2}{*}{Date\footnote{Start of observation (post-reduction).} (MJD TDB)} & Exposure & \multirow{2}{*}{Observation ID} & Name \\
& & time (ks)& & ([t]YYMMDD)\\ \hline
\renewcommand{\arraystretch}{1.5}
\multirow{2}{*}{\XMM\footnote{In all \XMM\ observations EPN detector was used.}} &53994.791448810  & 46.0 & 0404340101 & 060916\\
& 54000.527619667 & 29.2 & 0311792001&060922\\\hline
\multirow{5}{*}{\emph{Chandra}\footnote{In all \emph{Chandra} data ACIS detector was used.}} & 54005.283617719 & 15.7 & 6724 & c060927\\
& 54009.985545836 & 20.7 & 6725 & c061002\\
& 54017.265934068 & 26.2 & 6726 & c061009 \\
& 54036.293110632 & 15.7 & 8455 & c061028\\ 
& 54133.801451742 & 20.6 & 8506 & c070113\\ 
\hline
\multirow{5}{*}{\XMM$^b$} & 54148.378135941 & 17.6 & 0410580601& 070217\\
& 54331.412027612 & 23.7 & 0505290201& 070819\\
& 54511.305592759 & 29.8 & 0505290301& 080215\\
& 54698.508875675 & 30.7 & 0555350101& 080820\\
& 55067.333418577 & 41.4 & 0604380101& 090824\\
\hline
\multirow{2}{*}{\emph{Swift/PC}\footnote{In all \emph{Swift} observations we refer to the XRT}} & 55823.887347023 & 3.1 & 00030806020 & \\
& 55829.233491897 & 4.3 & 00030806022  & \\ \hline
\XMM$^b$ &  55831.936336540 & 16.7& 0679380501& 110927\\ \hline
\multirow{5}{*}{\emph{Swift/PC}$^d$} &  55835.185790895 & 3.7 & 00030806023 & \\
&  55839.066673333 &  3.7  & 00030806024  & \\
& 55842.093900977  &  3.9  & 00030806025 &\\
& 55844.092375078  &  4.0  & 00030806026 & \\
& 55849.040224474  &  8.8  & 00030806027 & \\ 
 \hline
\emph{Chandra}$^c$  & 55857.646333201   & 19.1 & 14360 & c111023 \\ \hline
\multirow{2}{*}{\emph{Swift/PC}$^d$}  & 55974.223899273 & 0.6 & 00030806028-29 \\
& 56001.015943907	& 2.4 & 00030806031 \\ \hline 
\hline
\end{tabular}
\end{minipage}
}
\caption{Summary of the observational data used in the paper}
\label{data}
\end{table*}

\begin{figure}
    \centering
      \includegraphics[width=.34\textwidth, angle=270]{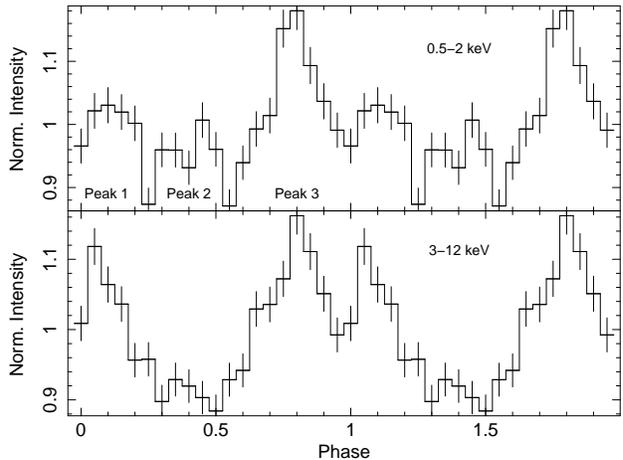}
    \caption{Upper panel: pulse profile of \src, observation 060922 at the $0.5 - 2.0$ keV energy range. 
      Lower panel: The same observation at the $3.0 - 12.0$ keV energy range. 
      Note that the peak at phase $\sim 0.4$ is missing at higher energies.} 
    \label{ene}
\end{figure}

\subsection{CXOU J164710.2-455216}
The source, CXOU J1647-45 for short, was discovered by Muno et al.
(2006) using \chandra\ observations, with a period of 10.6107(1) s.
An important feature of \cxou\ is that it very likely belongs to the young, 
massive Galactic starburst cluster Westerlund 1. This provides
hints about its progenitor and also about its distance.
Indeed, studies of the massive stellar population of Westerlund 1
indicate a distance of $\sim 5.0$ kpc and a progenitor with an initial
mass $M_i>40\:M_{\odot}$ (Crowther et al. 2006; Muno et al. 2006
Negueruela Clark \& Ritchie 2010).

Another prominent feature of \cxou, as mentioned before, is that it 
underwent an outburst with one of the largest flux enhancement observed 
up to now among the magnetars. On September 2006 the Burst Alert Telescope
(BAT) on board the \swift\ satellite detected
an intense burst in the direction of the Westerlund 1.
A second observation, performed  13 h later by \swift, with its narrow field 
instrument, the X-ray telescope (XRT), 
found \cxou\ brighter  by a factor of $\sim 300$. 
Between February 2007 and August 2009 we requested and obtained 
five  \emph{XMM-Newton}
pointings which, together with the September 2006 post-outburst
observations,  were aimed at studying the evolution of the timing and 
spectral properties of \cxou\ over a range covering a  
factor of about 50 in flux, from $\sim 10^{35}$ \ergs\ down to near the
quiescent level,  at a few $10^{33}$ \ergs\ (Campana \& Israel 2006). 
Deep observational campaign in the radio, near-infrared and hard X-ray 
bands did not detect any convincing counterpart (Muno et al. 2006; Israel et
al. 2007), in contrast with the results obtained for other transient magnetars, 
e.g. XTE J1810 (Israel et al. 2004; Camilo et al. 2006) and 1E 1547 
(Camilo et al. 2007; Israel et al. 2009).

On 2011 September 19 \swift-BAT recorded four relatively bright bursts 
from a position consistent with that of \cxou\ (Baumgartner et al. 2011), 
approximately five years after the 2006 outburst onset. A subsequent 
\swift-XRT pointing found  \cxou\ at a flux level of $\sim 7.8 \times 10^{-11}$ \ergscms, 
more than 200 times higher than its quiescent level 
($2.7 \times 10^{-13}$ \ergscms, Muno et al. 2007), and more than 100
times brighter than the latest \emph{XMM-Newton} pointing of 
August 2009 (Israel, Esposito \& Rea 2011): the pulsar entered a new
outburst phase. 

Several \swift\ observations were requested together with two director's 
discretionary time observations, one with \emph{XMM-Newton} and one with
\emph{Chandra}. The latter two were carried out 9 and 34 days after the
BAT trigger, respectively. A further  \emph{XMM-Newton} pointing 
scheduled for April 2012 was cancelled because of a strong solar storm.
The \emph{XMM-Newton} and  \emph{Chandra} pointings aimed at comparing 
the properties of the 2006 and 2011 outbursts.

\cxou\ 2006 outburst has been analyzed in previous 
investigations. In particular timing and spectral analyses have been 
performed by Israel et al. (2007) and Woods et al. (2011; both
phase-coherent), and An et al. (2013; period evolution).
Their timing results are summarized in Table \ref{altime}.
The phase-averaged fluxes and periods for
the 2011 outburst, as derived from the \emph{XMM-Newton} and  
\emph{Chandra} pointings, were reported 
by An et al. (2013). A detailed spectral and timing analysis is first 
reported in this paper where we present an extended, phase coherent 
long-term timing solution and phase-resolved spectroscopic analysis for 
both outbursts.
The implications, within the magnetar scenario, are also discussed by
means of state-of-the-art magnetothermal evolution simulations.

\section{OBSERVATIONS AND DATA PROCESSING}
For the 2006 outburst analysis we used data from eight \emph{XMM-Newton} and five 
\emph{Chandra} observations. For the 2011 outburst  one \emph{XMM-Newton}, one  
\emph{Chandra} and nine \emph{Swift} observations were used. A detailed log of 
all the collected  data can be found in Table \ref{data}.\\

The data reduction were performed following standard procedures 
and consisted of initial raw data calibration; filtering, including 
from solar flares and soft photons falres; correcting the photon's 
arrival times to the barycenter of the Solar system; source and 
background extraction; pileup checks; and  spectral data rebining 
and oversampling (see Section \ref{spec} for details). 
The reduction procedures were performed using the official Science Analysis System (\textsc{sas}) package 
(version 12.0.1 release: \textsc{xmmsas\_20110223\_1801-11.0.0})
for the \XMM\ data, and the Chandra Interactive Analysis of Observations 
(\textsc{ciao}) system (version 4.4) for the \emph{Chandra} data. 
The \swift\ data were processed and filtered with standard procedures 
and quality cuts\footnote{See http://swift.gsfc.nasa.gov/docs/swift/analysis/ 
for more details.} using \textsc{ftools} tasks in the 
\textsc{heasoft} software package (v.6.12) and the calibration 
files in the 2012-02-06 \textsc{caldb} release.  

For the spectral analysis we used \textsc{xspec} (version 12.7.1) 
and for the timing, \textsc{xronos} (version 5.22) and pipelines developed
in-house for the phase-fitting procedures.

\section{TIMING}

\subsection{2006 -- 2009 Outburst}
\label{timing2006}

\begin{figure}
  \centering
  \includegraphics[width=0.83\textwidth, angle=-90]{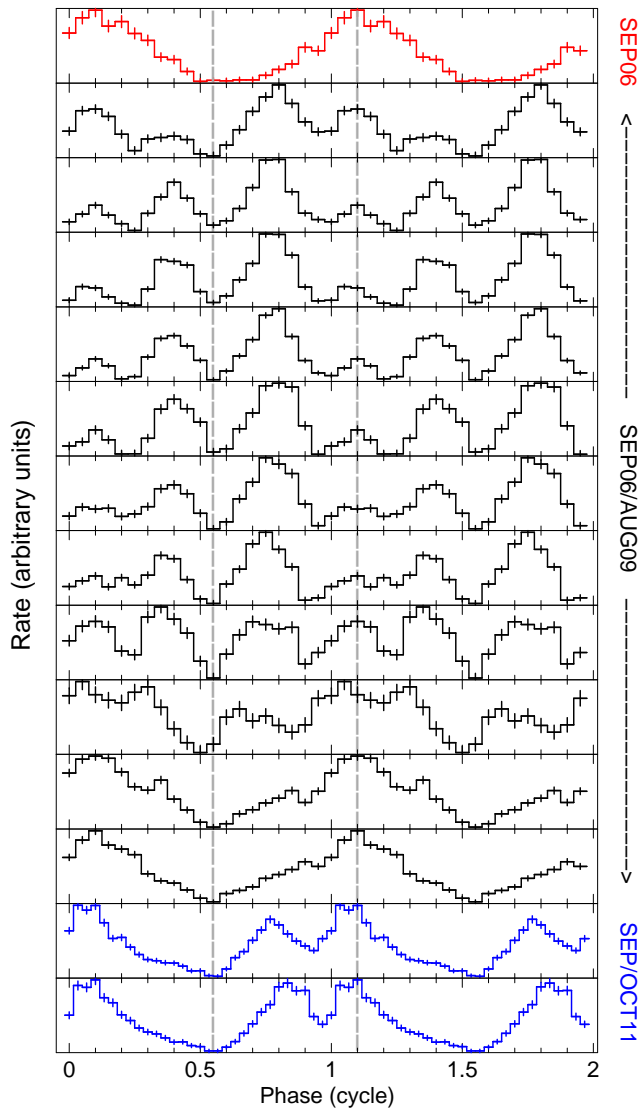}
  \caption{Pulse shape evolution over time. \emph{XMM-Newton} 2006 pre-outburst 
    observation (red line), \emph{XMM-Newton} and \emph{Chandra} data during 
    2006 outburst by using our phase-coherent timing solution (black lines), 
    and \emph{XMM-Newton} and \emph{Chandra} data during 2011 outburst (blue lines). 
    Both the 2006 pre-outburst and  2011 outburst folded light curves have 
    been shifted in phase in order to align their minima with those of the 
    2006 outburst. For better  visualization the data has been normalized 
    to the average intensity, the pulse fraction values and its evolution 
    are shown in Fig. \ref{PF}.}
  \label{timing}
\end{figure}

For the timing analysis of the 2006 -- 2009 outburst, 
we used all the data (see Table \ref{data}) between 060922  
and 090824.

With  the only exception of 090824, all the pulse profiles used in the 
timing analysis present a three-peaked structure, 
the relative amplitudes and phases of the peaks were such 
that it was not straightforward to unambiguously follow the signal 
phase evolution throughout the outburst decay.  To pinpoint 
the correct signal phase evolution 
we combined information from the peaks relative (phase) positions 
and spectral data. For instance, 
the first peak (see reference on Fig. \ref{ene}) tends to be wider 
at higher energies ($> 3$ keV) while the second one is 
significantly weaker at low ($< 2$ keV) energies 
(See for example Fig. \ref{ene}; see also Muno et al. 2007).  

Based on the above findings we were able to track the peak 
correspondence for the whole time span from September 2006 
to August 2009 (see Fig. \ref{timing}).  We note that during the latter observation  
 the pulse shape had almost returned to the quiescence single-peak 
profile, while at the beginning of the 2011 outburst the pulses 
showed again a multiple-peak structure compatible with that of 
the first observations of the 2006 outburst, though with a 
rather larger pulsed fraction (see below).

To obtain a phase-coherent timing solution, we started by folding 
the data into 20 bins per cycle. We considered only events in the  $0.5 - 4.0$ keV energy  interval, 
since in some observations the shape of the pulse slightly changes at higher energies 
(see e.g. Fig. \ref{ene}), which could affect the phase-fitting procedure. 
We started by dividing the first observation in four segments folded at the period 
and period derivative reported by Woods et al. 2011 (quadratic fit). Next, 
we performed the phase-fit procedure obtaining a new
solution and repeated the procedure iteratively using the new solution 
and including the subsequent observation. For details on the phase-fitting 
procedure see Dall'Osso et al. (2003).

\begin{figure*}
  \begin{minipage}{177mm}
    \centering
    \includegraphics[width=1.01\textwidth]{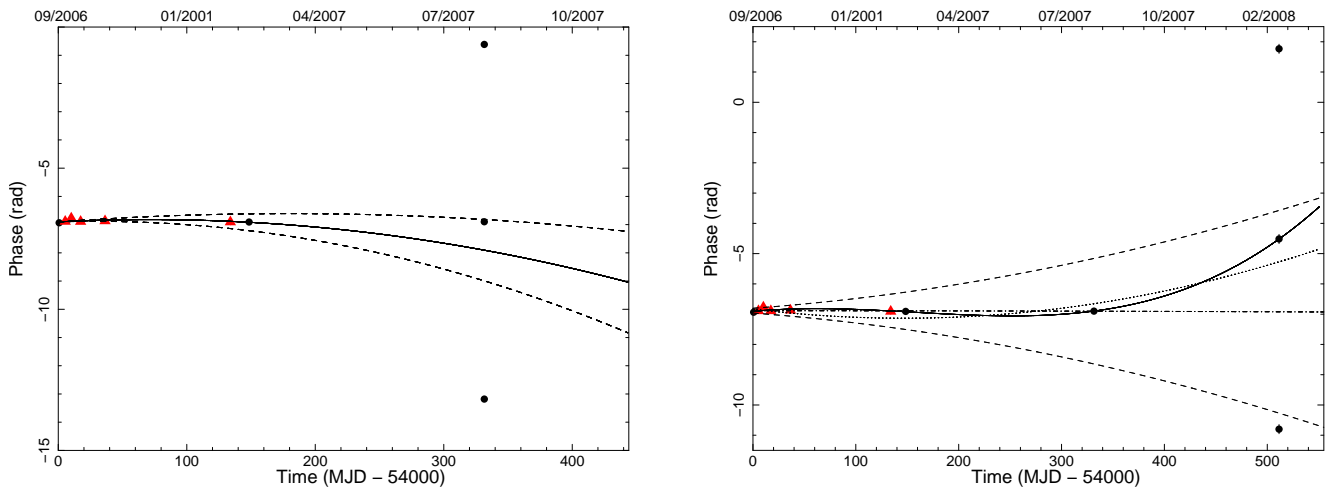}
    \caption{Left panel: phase connected observations up to Feb. 2007 ($\sim 148 $ d) using only $P$ and $\dot{P}$ terms. 
      A quadratic fit (solid line) and a $3\sigma$ area of its parameters (delimited by dashed lines) are shown. 
      It defines the area where there is a $3\sigma$ certainty of phase coherence. For the next observation, 
      Aug. 2007 ($\sim 331 $ d), one and only one cycle falls inside the delimited area, thus phase coherence 
      is maintained at a 3$\sigma$ confidence level and we proceed to correct and extend our timing solution. 
      Right panel: the dashed-dot line represents the timing solution up to Aug. 2007 ($\sim 331 $ d) with only 
      $P$ and $\dot{P}$ terms. Correspondingly, the cone delimited by dashed lines represent the area where there 
      is a $3\sigma$ certainty of phase coherence for that solution. Solid line: subsequent cubic 
      ($P$, $\dot{P}$ and $\ddot{P}$) timing solution. Dotted line: best quadratic fit (black circles are 
      \XMM\ data, red triangles represent \chandra\ observations).} 
    \label{311}
  \end{minipage}
\end{figure*}

Up to August 2007 (Obs. 070819) a period ($P$) and period derivative ($\dot{P}$)
components were enough to obtain a phase-coherent timing solution
(see Fig. \ref{311}, Left panel). After that point a $\ddot{P}$ component becomes necessary 
(F test at $\sim 4\sigma$ (99.992\%) confidence level (c.l.), see Fig. \ref{311} Right panel). 

It can be seen in Fig. \ref{311}, left panel, that at this 
epoch the phase connection is maintained marginally at  
$3\sigma$, (at $5\sigma$ the phase coherency is lost and there is a two-cycles ambiguity).
While the whole phase-fitting process was performed at a $3\sigma$ c.l., 
in this marginal case we performed an additional test: we separately 
assumed each of the possible (at $5\sigma$ c.l.) cycles as correct 
and continued parallelly the phase-fitting iteration with the next 
observation(s), obtaining two different phase evolution tracks. 
Then, we look if any of the phase track works well without the addition 
of any further, higher order components, other than those already present 
in the solution (constant, linear, quadratic and cubic terms).  
We found that actually only one track yielded a feasible solution, 
and it coincided with the one found initially at the $3\sigma$ c.l., see Fig. \ref{311}, right panel.

\begin{figure}
  \centering
  \includegraphics[width=0.35\textwidth, angle=270]{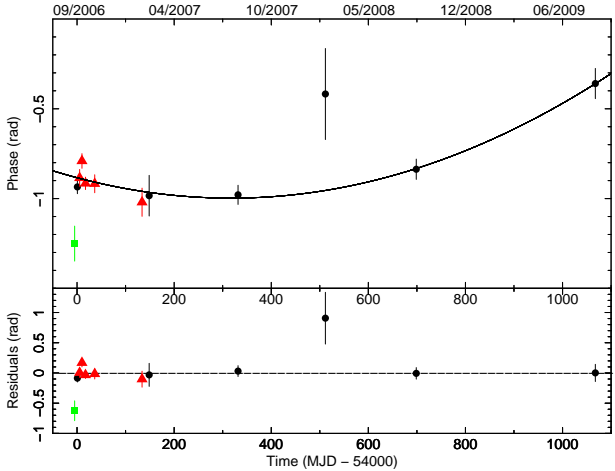}
  \caption{Upper panel: Phases of the \XMM\ (black circles) and \chandra\ (red triangles) observations of \src\ minus a cubic component. 
    The solid line represents the final timing solution. Lower panel: fit residuals. The green square represents the pre-outburst \XMM\ observation (060916).}
  \label{timf}
\end{figure}

For all the observations in the 2006 September 22 - 2009 August 24 
time interval (Fig. \ref{timf}) the resulting phase-coherent 
solution based on phase residual versus time fits gave a best-fitting
P= 10.61065583(4) s, $\dot{P}  = 9.72(1) \times 10^{-13}$ s/s and 
$\ddot{P} = -1.05(5)\times10^{-20}$ s s$^{-2}$  (all with  $1\sigma$ uncertainty) 
and MJD 54008.0 as reference epoch, see Table \ref{altime}. 
Our solution shows a good consistency with the data, for instance the $\chi^2$/dof for the whole set of data
(\XMM\ and \chandra) is $8.95/8$ (see Fig. \ref{timf}).

Note that since we focused on a long-term timing solutions, 
the reported glitch (Israel et al. 2007) near the outburst 
onset characterized by a short recovery time of $\sim 1$ week, 
does not affect our solution. 
Earlier works have looked into it, see Israel et al. (2007) 
and Woods et al. (2011); the detailed short-time analysis 
required to look into it is outside the scope of this paper.
We only note that the extrapolation of the above reported 
phase-coherent solution backwards to the first pre-outburst 
observation implies a $\Delta \phi$ of $\sim 0.06$ cycles, 
and a $\Delta\nu/\nu \sim 1.8(6) \times 10^{-5}$, 
in agreement with an upper limit for $\Delta\nu/\nu < 1.5 \times 10^{-5}$ 
reported by Woods et al. (2011).

\begin{table*}
\begin{minipage}[H]{150mm}
\renewcommand{\arraystretch}{1.3}
\resizebox{12cm}{!} {
\begin{minipage}[H]{12cm}
\begin{tabular}[H]{|l|l|l|c|c|c|l}
\hline
\multicolumn{7}{|c|}{Summary of CXOU J164710.2-455216 timing solutions for the 2006-2009 outburst decay} \\
\hline
\renewcommand{\arraystretch}{1.0}
 & Epoch&Period &$\dot{P}$ \footnote{Period time derivative}&$\ddot{P}$ \footnote{Second period time derivative}&B$\times 10 ^{14}$ \footnote{$P-\dot{P}$ estimated surface dipolar magnetic field at reported epoch, assuming an orthogonally rotating neutron star of radius 10\,km and moment of inertia $10^{45}$ g cm$^2$.}&Notes\\
& (MJD) & (s) & ($10^{-12 }$ s/s) & ($ 10^{-20}$  s s$^{-2}$) & (Gauss) &
\\\hline
\renewcommand{\arraystretch}{1.5}
Israel et al. 2007     &53999.0& 10.6106549(2) & 0.92(4)   & -      &  1.0  & Coherent, Quadratic fit\\ 
Woods et al. 2010      &54008.0& 10.6106563(1) & 0.83(2)   & -      &  0.95 & Coherent, Quadratic fit\\ 
Woods et al. 2010      &54008.0& 10.6106558(2) & 1.3(1)    & -10(1) &  1.21\footnote{Instant value at reported epoch. Note that in Woods et al. (2011) only the average value over the time span of their analysis is reported ($0.89 \times 10^{14}$  G).} &  Coherent, Cubic fit\\          
An et al. 2013         &53999.1& 10.61064(2)   & $<$0.4(6) & -      & $<$ 0.7& Non-coherent, linear fit \\ 
This work              &54008.0& 10.61065583(4)& 0.972(1)  &-1.05(5)&  1.04 & Coherent, Cubic fit\\        
\hline
\end{tabular}
\end{minipage}
}
\caption{Summary of previous timing solutions}
\label{altime}
\end{minipage}
\end{table*}

\begin{figure}
  \centering
  \includegraphics[width=0.97\columnwidth, angle=270]{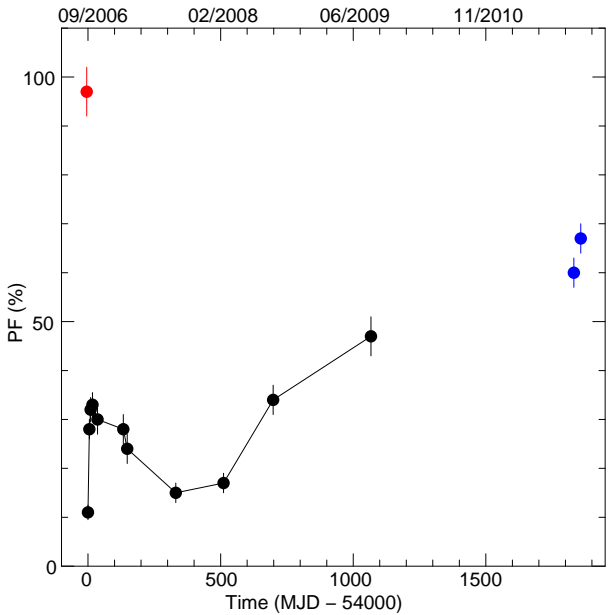}
  \caption{Evolution of the pulsed fraction (see the text for definition) as a function of time (in  truncated JD): black filled circles mark the 2006-2009 \cxou\ outburst, while the pre-outburst and 2011 outburst values are reported in red and blue, respectively.}
  \label{PF}
\end{figure}

We also studied the evolution  of the pulsed fraction, defined as 
the semi-amplitude of the sinusoid divided by the average count rate. 
Because of the complexity of the pulse shape, in many cases three sinusoids 
are needed in order to well reproduce the profiles. We inferred the 
square root of the quadratic sums of the semi-amplitude of each sinusoid 
as a rather better estimate of the profile pulsed fraction. The latter 
quantity is shown in Fig.\,\ref{PF} as a function of time since the 
first, pre-outburst \XMM\ pointing.

\subsection{2011 Outburst}
\label{timing2011}
We started by inferring an accurate $P$ measurement for the \XMM\ observation 
110927 by means of a phase-fitting algorithm similarly to the approach adopted for the 2006 outburst data.  
We found a best period of $P=10.61066(1)$ s. The relative accuracy was enough to phase-connect further data sets.
  
The  relative phases  and amplitudes are such that the signal
phase evolution could  be followed unambiguously for the \emph{Swift} 
and \emph{Chandra}  observations carried out during the 19 September 
2011 - 23 October 2011 time interval (see latest two folded light curves 
in Fig. \ref{timing} for  \XMM\ and \emph{Chandra}). Within this interval 
a term taking into account for the period evolution  started to be 
statistically needed. By adding a quadratic component to the phase history 
we obtained a best-fitting period of $P=10.610673(2)$ s  and $\dot{P} = 
3.5(1.0) \times 10^{-12}$ s s$^{-1}$ (1$\sigma$ uncertainties are reported;  
epoch = 55823.0 MJD; $\chi^2$/dof = $11/7$). The subsequent source seasonal 
visibility window opened in February 2012 during which two further 
\emph{Swift} pointings were carried out. Unfortunately a relatively long 
\XMM\ pointing, scheduled on March 2012, was deleted due to bad space weather 
(intense solar storm). We therefore used the remaining two low-statistics 
\emph{Swift} pointings in order to further refine the 2011 timing solution. 

Unfortunately, the 2011 timing solution accuracy was not good enough to keep 
unambiguously the coherence until  the 2012   \emph{Swift} pointings, and as 
a consequence three different solutions become, therefore, possible  (starting 
from low $\dot{P}$ to larger values): (a) $P=10.6106787(4)$ s  and 
$\dot{P} = 7(1) \times 10^{-13}$ s s$^{-1}$ ($\chi^2$/dof = $19/9$), (b) 
$P=10.6106761(4)$ s  and $\dot{P} = 2.2(1) \times 10^{-12}$ s s$^{-1}$ 
($\chi^2$/dof = $19/9$), and (c) $P=10.6106723(4)$ s  and 
$\dot{P} = 4.3(1) \times 10^{-12}$ s s$^{-1}$ ($\chi^2$/dof = $15/9$). 
We note that solution (a) is in agreement, within uncertainties, with 
the 2006 outburst timing parameters, while solution (c), with a slightly better 
reduced $\chi^2$, correspond to a rather high $\dot{P}$. Moreover, solutions (b) 
and (c) are within 2$\sigma$ from the 2011 timing solution, while solution (a) 
is slightly farther than 3$\sigma$.

During the 2011 outburst pulse profiles returns to a multiple-peak configuration, and the pulse fractions are in between those measured for the 2006 pre-outburst observation and for the 2006-2009 outburst  (see also Figs.\,\ref{timing} and \ref{PF}), with the 2011 values being on the extrapolation of the 2008-2009 trend. 

\section{SPECTRAL ANALYSIS}
\label{spec}
\subsection{2006 -- 2009 Outburst}
\label{spec2006}

For the spectral analysis we used only the \emph{XMM-Newton} observations 
in order to rely upon higher statistics data and the same instrument 
(therefore minimizing the possible intercalibration issues among 
different detectors).  
We applied our final timing solution to the data and we performed a 
pulse phase spectroscopy (PPS) over the whole time interval 
of validity of the timing solution. Since the pulse profile displays
such  a complex multipeak structure, where each peak seems to evolve 
differently (from the point of view of their relative fluxes, see Fig. 
\ref{timing}) over time, 
it was important to study each different component (such as peaks and
minima) separately, as a function of time and to find out if there were  
any spectral peculiarities along the pulse phase. \\

\begin{figure}
  \centering
  \includegraphics[width=0.33\textwidth, angle=270]{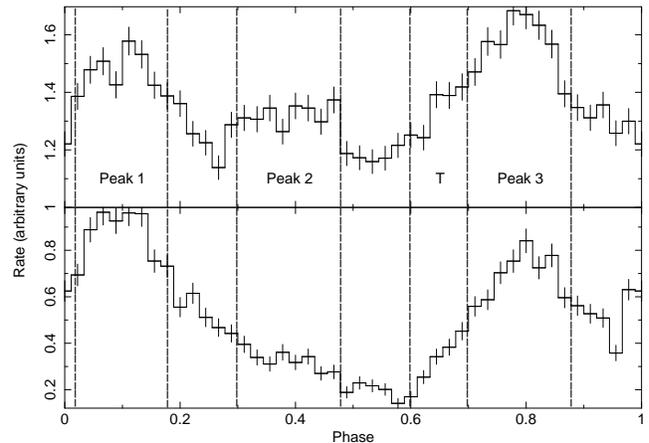}
  \caption{Pulse phase intervals used in the phase-resolved spectroscopy. 
    The phase zero-point is the same as in Fig. \ref{timing}. 
    The letter ``T'' denotes the transition region (see text for details). 
    Upper panel: first \XMM\ observation of the 2006--2009 outburst (060916).
    Lower panel: first \XMM\ observation of the 2011 outburst (110927).}
  \label{pps1}
\end{figure}

With this aim in mind we first divided the pulse phase in seven parts: three peaks, 
three minima and a transition region which shows different spectral characteristics
in high (3--12 keV) and low (0.5--2 keV) energies, with respect to those of the peak 
adjacent to it (peak 3; see the peaks nomenclature and spectral bins used 
in the PPS in Fig. \ref{pps1}). 
This pulse-profile segmentation allows us to follow the evolution of 
the peaks and the minima with the maximum possible signal-to-noise ratio.\\

To obtain the phase-resolved spectra, first we used `\verb phasecalc '\footnote
{Part of the Science Analysis System (\textsc{sas}) package 
(used version 12.0.1 release: \textsc{xmmsas\_20110223\_1801-11.0.0})\label{sas}}
to calculate the phases on each observation events file.
Then used `\verb tabgtigen '\footref{sas} and `\verb evselect '\footref{sas} to obtain 
the seven event files per observation filtering by phases and subsequently
extract their spectra. `\verb arfgen ' and `\verb rmfgen ' were used to generate the 
ancillary response files and the redistribution matrix files, respectively.
Then we used `\verb grppha '\footnote{Part of HEASoft (used heasoft-6.12)\label{hea}}
to rebin the spectra to ensure that each spectral channel had at least 30
counts and to oversample the instrumental energy resolution by a factor of three.

Previous spectral analysis on this source has been performed by 
Israel et al. 2007; Woods et al. 2011 and An et al. 2013. 
In all previous works the average spectra was fitted with 
an absorbed, single blackbody (BB) plus a Power Law (PL). 
However, Albano et al. (2010) based their analysis on a 
more physical model taking into account the effect of the 
magnetosphere twist. Physical and geometrical parameters 
were recovered from the joint modeling of the pulse profiles 
and spectra. The resulting best-fits for the light curves,
consisting of three NTZang spectra (Nobili, Turolla \& Zane 2008)
were then used to fit the observed spectra, mimicking the 
magnetospheric reprocessing of photons from three regions 
of the NS surface at different 
temperatures, obtaining good agreement with the data. 
(See Albano et al. 2010 for details).
We tried a similar spectral decomposition, but due to the 
relatively high number of free parameters in the latter model 
and the lower statistics of phase-resolved spectra resulted 
in a reduced $\chi^2$ systematically lower than 1. Therefore, 
we decided to use the 
closest possible model to that used by Albano et al. (2010) 
by assuming a three absorbed BB components: 
\verb|phabs(bbodyrad1 + bbodyrad2 + bbodyrad3)| in \verb XSPEC \footref{hea}.  
One of the BBs had a fixed temperature of 150 eV which is meant 
to correspond to the ``cool'' fraction of the NS surface; and the other two BBs 
to a hotspot and a warm zone around it (or, in principle, any 
other two-temperature configuration), and were left free to vary 
between observations. Representing a thermal map of the whole NS
surface, their temperatures, where forced to be the same for 
all the phase intervals, in each epoch (see Albano et al. 2010 
for more details on the geometric model).
The absorbing column density was fixed to $2.4 \times 10^{22}$ cm$^{-2}$, 
based on the phase average spectral fits. 
Such a configuration, with an appropriate spin and magnetic axis 
and line-of-sight angles, may reproduce the three-peaked pulse 
profile (see Albano et al. 2010). 

\begin{figure}
  \centering
  \includegraphics[width=0.37\textwidth, angle=270]{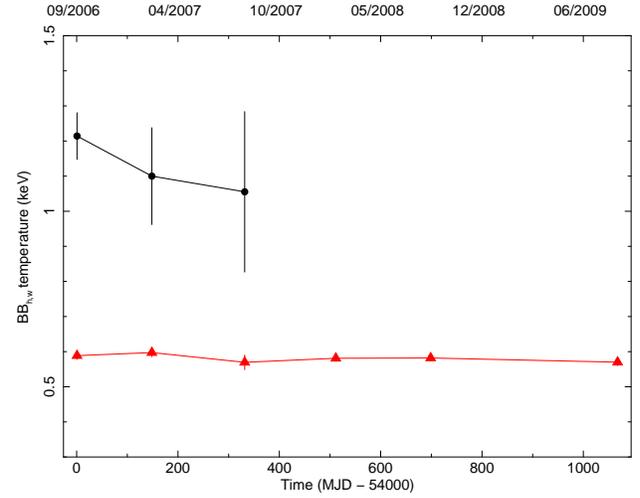}
  \caption{Temperature evolution of the BB used for modeling the peaks spectra. 
    Black circles correspond to the hard component, red triangles to the warm component 
    The zero point in time represents the onset of the 2006 outburst.
  }
  \label{pps2}
\end{figure}

\begin{figure}
  \centering
  \includegraphics[width=0.49\textwidth, angle=0]{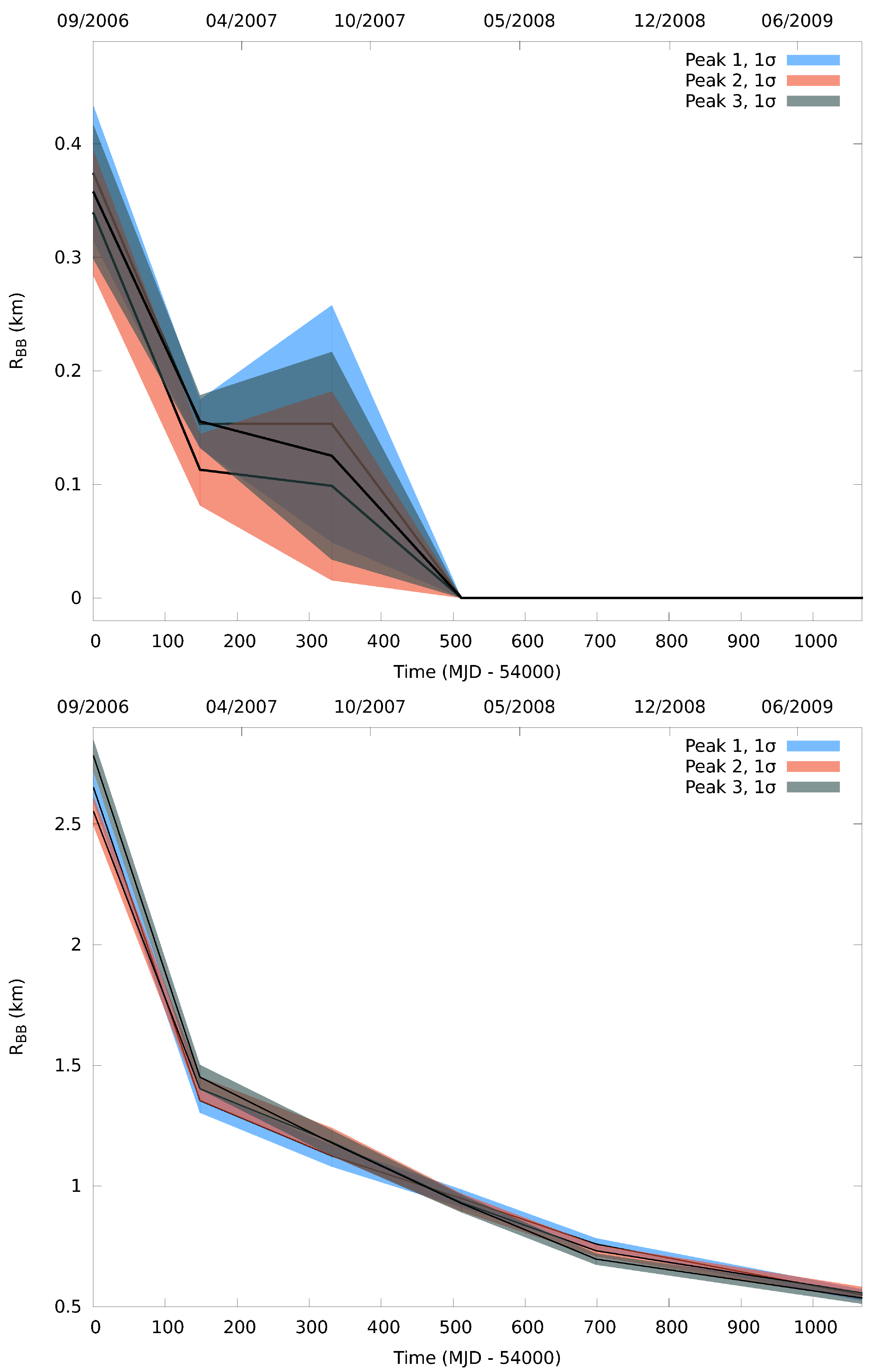}
  \caption{Evolution of the blackbodies radii ($R_{BB}$) used for modeling the peaks spectra. Upper panel: 
    hard component; lower panel: warm component. 
  }
  \label{pps2b}
\end{figure}

\begin{figure*}
  \centering
  \includegraphics[width=0.95\textwidth, angle=0]{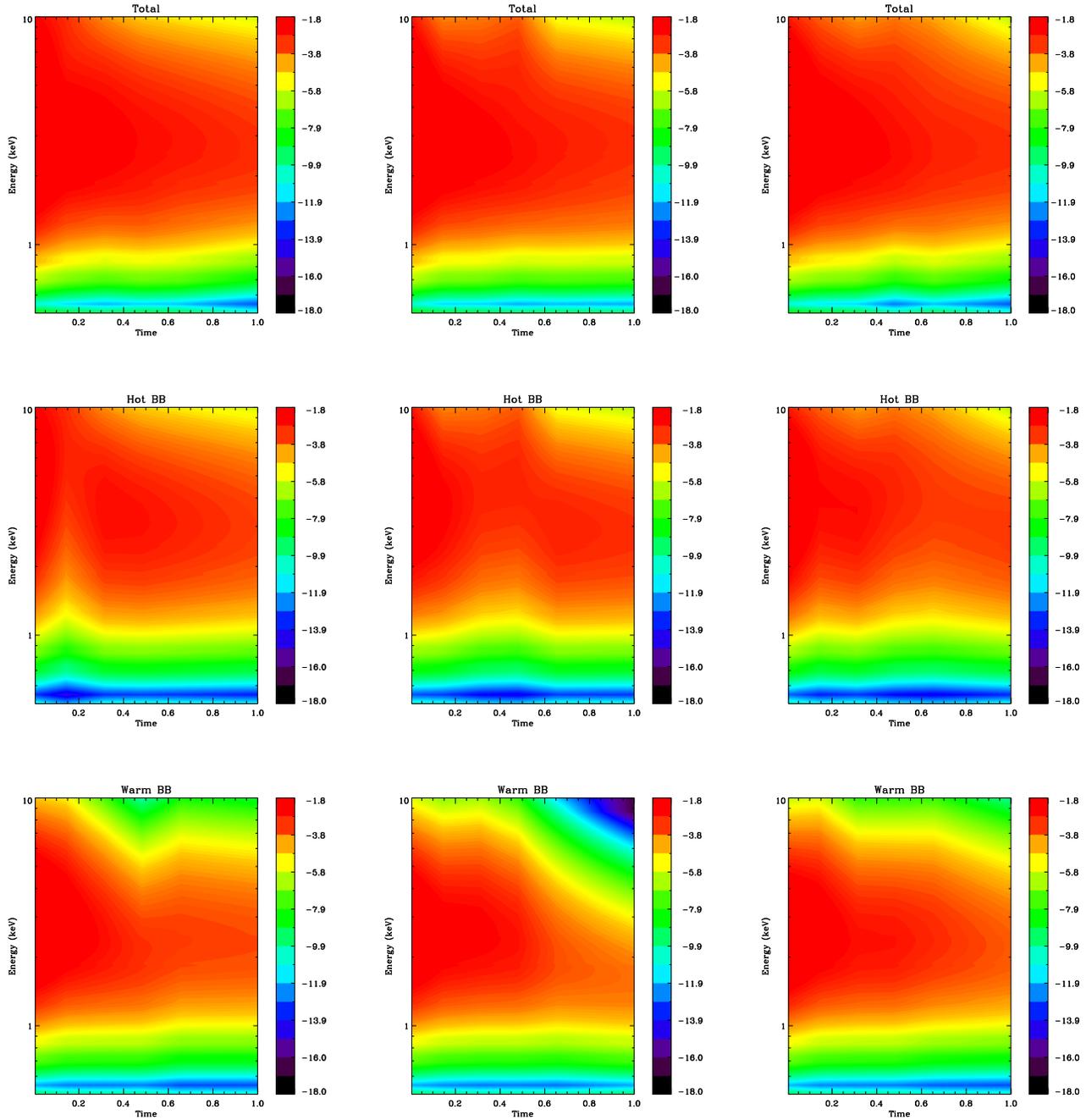}
  \caption{Dynamic spectral profiles. Energy-resolved Flux (colors) evolution 
    over the duration of the 2006 outburst campaign for the three pulse-profile 
    peaks. Each column corresponds to one of the peaks: first (left), second (center) 
    and third (right). The three rows represent in the time/energy plane the contour 
    plots for the total (upper), hot BB (middle) and warm BB (lower) $\nu F \nu$ fluxes. 
    The color scale is in log units of keV$^{2}$ (photons cm$^{-2}$ s$^{-1}$ keV$^{-1}$). 
    The x-axis is time MJD - 54000, normalized to the duration of the 2006 outburst campaign 
    ($\sim 1067.5$ d).
  }
  \label{pps3}
\end{figure*}

In several cases the statistics of the minima's spectra was not 
good enough to obtain an acceptable spectral fit, being overfited 
by our 1(kT-fixed)+2BB model. Since, the problem of low statistics 
affected most of then, we decided to focus on the pulse-profile peaks. 
The resulting BB parameters are presented in Figs. \ref{pps2}, \ref{pps2b}
and Table \ref{pps}. In Fig. \ref{pps3} dynamic spectral profiles of each 
peak are presented, in the plots the flux density distribution over the 
1--10 keV energy range, over the 3 yr of the 2006 outburst-decay 
campaign.

\begin{figure}
  \centering
  \includegraphics[width=.45\textwidth, angle=0]{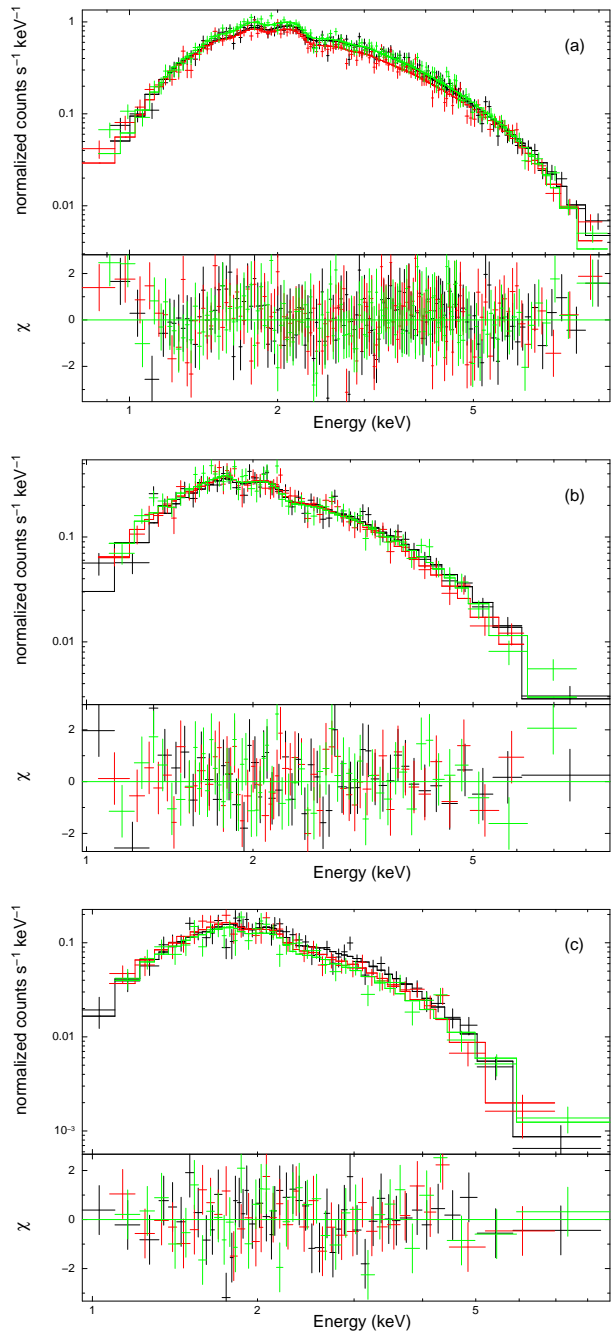}
  \caption{Pulse phase-resolved spectra of the peaks. 060922(a), 
  070819(b) and 080820(c). Peaks spectral relative evolution at a glance.
  First peak -- black; second peak -- red; third peak -- green. 
  See Fig. \ref{pps1} and \ref{timing} for peaks reference.}
  \label{peaksp}
\end{figure}

\begin{figure}
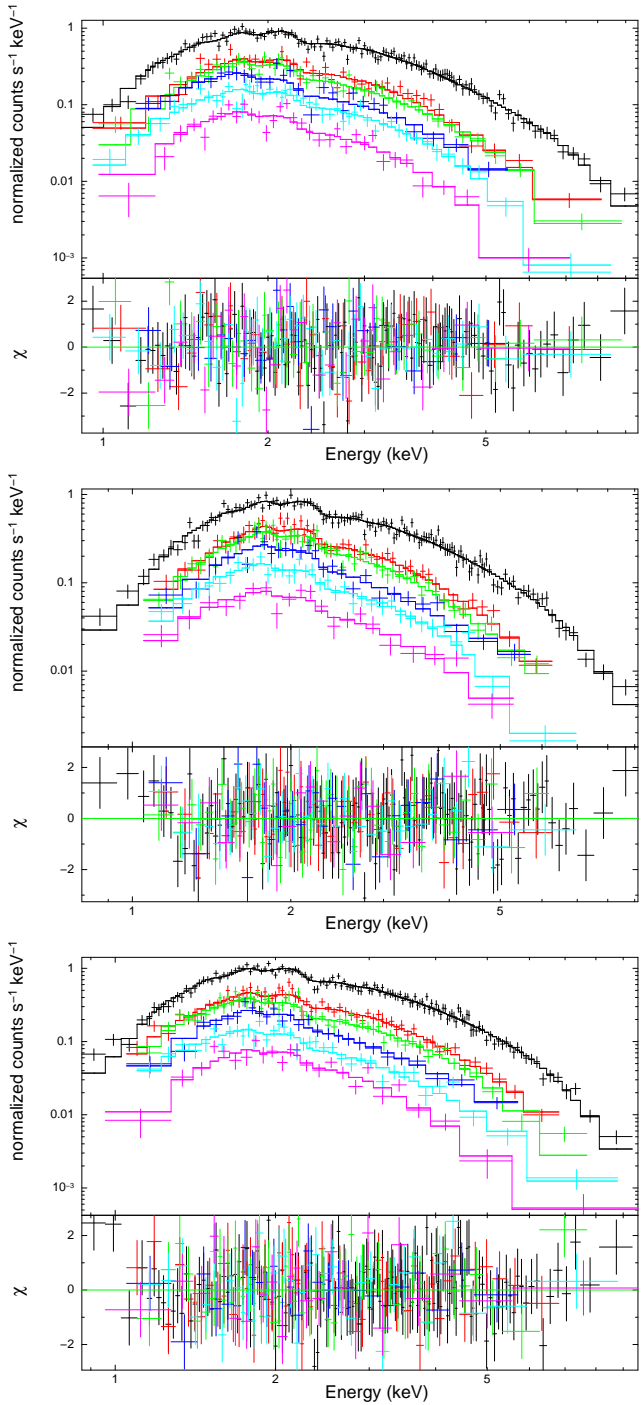

  \centering
  \includegraphics[width=0.35\textwidth, angle=270]{fig8UP.ps}
  \includegraphics[width=0.35\textwidth, angle=270]{fig8MID.ps}
  \includegraphics[width=0.35\textwidth, angle=270]{fig8LOW.ps}
  \caption{Phase resolved spectral evolution of \src. 
    The solid lines represent the 1+2BB absorbed model for the 
    peak 1 (upper panel), peak 2 (middle panel) and peak 3 (lower panel),
    see Fig. \ref{pps1} and \ref{timing} for reference.
    Black: 060922; red: 070217; green: 070819; blue: 080215; 
    cyan: 080820; magenta: 090824. Residuals are shown in the lower part of each panel.}
  \label{peakssp}
\end{figure}

Our analysis indicates that, indeed, there are spectral differences between them; 
both on single observations and on their after-burst relaxing evolution over time
(see Figs. \ref{peaksp} and \ref{peakssp}). For instance, peak 2 (see reference 
on Fig. \ref{pps1}) is softer then the others, and peak 1 (which correspond to the 
quiescence peak), is harder. 
The temperatures of the BB do not vary significantly. Specially the warm component
shows a very steady value of $\sim 0.58$ keV. The hot component may be more
variable, but the errors do not allow us to draw concrete conclusions in this regard.
On the other hand, the evolution of the BB-radius shows a constant and significant 
shrinking of both components. Indeed, the hot component disappears about 500 d
after the outburst onset  
, see Figs. \ref{pps2}--\ref{pps3} and Table \ref{pps}.

\subsection{2011 Outburst}
\label{spec2011}
Using the same phase intervals as for the 2006 outburst data and the 2011 
timing solution, we performed a PPS for the first \XMM\ observation of 
the 2011 outburst. The phase intervals were 
extracted with the same methods and fitted with the same models used for 
the 2006-2009 data (see Section \ref{spec2006});
with the new 2011 timing solution and the same peaks identification scheme
as for the 2006-2009 data (see Section \ref{timing2006}). The results are reported 
in Table \ref{pps}.

\begin{table*}
\begin{minipage}{110mm}
\renewcommand{\arraystretch}{1.3}
\resizebox{8cm}{!} {
\begin{minipage}{7.5cm}
\begin{tabular}{|l|l|c|c|c|c|}
\hline
\multicolumn{6}{|c|}{\src\ 1+2-blackbody spectral fit} \\
\hline
Peak\footnote{See Figs. \ref{pps1} and \ref{timing} for reference.} & Obs. ID& T$_w$ (keV) & R$_{BB}^W$ (Km) & T$_H$ (keV) & R$_{BB}^H$ (Km)\\ \hline
\multirow{6}{*}{First}                   & 060922 & 0.59 $\pm$  0.01  & 2.65 $\pm$ 0.06  & 1.21  $\pm$  0.07    &  0.37  $\pm$  0.06  \\
                                         & 070217 & 0.60 $\pm$  0.01  & 1.35 $\pm$ 0.05  & 1.10  $\pm$  0.14    &  0.15  $\pm$  0.02  \\
                                         & 070819 & 0.57 $\pm$  0.02  & 1.12 $\pm$ 0.05  & 1.06  $\pm$  0.23    &  0.15  $\pm$  0.10  \\
                (quiescent)              & 080215 & 0.58 $\pm$  0.01  & 0.95 $\pm$ 0.04  &      --         &        --             \\
  $\chi^2_{red} = 1.1865$                 & 080820 & 0.58 $\pm$  0.01  & 0.76 $\pm$ 0.03  &     --          &        --             \\
                                         & 090824 & 0.57 $\pm$  0.01  & 0.55 $\pm$ 0.03  &     --          &         --            \\ \cline{2-6}
                                         & 110927 &  --  &     --       & 0.77  $\pm$  0.01    & 0.23  $\pm$  0.01  \\ \hline \hline
\multirow{6}{*}{Second}                  & 060922 & 0.59 $\pm$  0.01  & 2.55 $\pm$ 0.06  & 1.21  $\pm$  0.07    & 0.34  $\pm$  0.06  \\
                                         & 070217 & 0.60 $\pm$  0.01  & 1.40 $\pm$ 0.05  & 1.10  $\pm$  0.14    & 0.11  $\pm$  0.03  \\
                                         & 070819 & 0.57 $\pm$  0.02  & 1.18 $\pm$ 0.06  & 1.06  $\pm$  0.23    & 0.10  $\pm$  0.09  \\
                                         & 080215 & 0.58 $\pm$  0.01  & 0.93 $\pm$ 0.04  &      --       &        --            \\ 
         $\chi^2_{red} = 1.0522$          & 080820 & 0.58 $\pm$  0.01  & 0.73 $\pm$ 0.03  &     --        &        --            \\
                                         & 090824 & 0.57 $\pm$  0.01  & 0.56 $\pm$ 0.03  &      --       &        --            \\ \cline{2-6}
                                         & 110927 &   --  &  --   &  0.77  $\pm$  0.01   & 0.20  $\pm$  0.01  \\ \hline \hline
\multirow{6}{*}{Third}                   & 060922 & 0.59 $\pm$  0.01  & 2.78 $\pm$ 0.07  &  1.21  $\pm$  0.07   & 0.36  $\pm$  0.06  \\
                                         & 070217 & 0.60 $\pm$  0.01  & 1.45 $\pm$ 0.05  &  1.10  $\pm$  0.14   & 0.16  $\pm$  0.03  \\
                                         & 070819 & 0.57 $\pm$  0.02  & 1.18 $\pm$ 0.06  &  1.06  $\pm$  0.23   & 0.13  $\pm$  0.09  \\
                                         & 080215 & 0.58 $\pm$  0.01  & 0.93 $\pm$ 0.04  &     --       &         --           \\
         $\chi^2_{red} = 1.0516$          & 080820 & 0.58 $\pm$  0.01  & 0.70 $\pm$ 0.03  &     --       &         --           \\ 
                                         & 090824 & 0.57 $\pm$  0.01  & 0.54 $\pm$ 0.03  &      --      &          --          \\ \cline{2-6}
                                         & 110927 &  --  &  --  &   0.77 $\pm$ 0.01    & 0.23  $\pm$  0.01  \\ \hline \hline
\hline
\end{tabular}
\end{minipage}
}
\caption{1+2BB fit spectral parameters. Obs 110927 corresponds to the successive outburst. See the text for details. }
\label{pps}
\end{minipage}
\end{table*}

\section{DISCUSSION}
\label{discussion}
\subsection{Timing}

Significant changes in the pulse profile during the outburst decay 
mean that peaks identification and the way of taking into account 
their variations in relative phase (among peaks), intensity and 
shape, is important in order to successfully phase connect the 
observations. For instance Woods et al. (2011) cite the ``extreme 
change in pulse profile'' as the reason why they were not able to 
phase connect the 070819 observation with their coherent solution. 
On the other hand, An et al. (2013) cite a large time separation
between 070819 and the previous observation as the cause of their 
phase connection loss. As mentioned before (see Fig. \ref{ene}) at 
different energy ranges the peaks behave differently. This fact, 
coupled with measurements of the relative phase distances between 
peaks allowed us to identify them. Once we obtained a consistent 
peak identification, we had no problems to keep the phase coherence, 
see Fig. \ref{311}, left panel. We believe that discrepancies with 
respect to previous published results may be due to the different 
assumptions used for the phase-fitting algorithm. 
 
The new spin-down value $\dot{P}̇ \simeq 9.7 \times 10^{-13}\;$s s$^{-1}$ 
is similar to that of the two previous 
$P$ and $\dot{P}$ 
solutions: $\dot{P} \simeq 9.2 \times 10^{-13}\;$s s$^{-1}$ 
derived by Israel et al. (2007) and $\dot{P} \simeq 8.3 \times 10^{-13}\;$s s$^{-1}$ 
reported by Woods et al. (2011) but significantly smaller 
then the one of the cubic solution of Woods et al. 2011 
($\dot{P} \simeq 13 \times 10^{-13}\;$s s$^{-1}$), who considered a 
shorter data sample spanning from 2006 September 23 to 2007 February 17. 
This may be due to a decrease of the spin-down rate throughout the outburst decay.

The $P$ and $\dot{P}̇$ values inferred imply a surface dipolar field 
B $\sim 1.0 \times 10 ^{14}$ G using the conventional formula at the 
equator B $= 3.2 \times 10 ^{19} (P\dot{P}̇)^{-1/2}$, assuming an 
orthogonally rotating neutron star of radius 10\,km and moment of 
inertia 10$^{45}$ g cm$^2$. This estimate lays in the standard magnetar 
range and agrees with the magnetar nature of \src. 

\subsection{Outburst decay}

In previous (phase-averaged) studies, spectra have been 
analyzed during the outburst decay and fitted with an absorbed
PL plus a BB (Woods et al. 2011; An. et al. 2013).
In those works the BB evolution during the outburst agrees with that 
of the general trend of the warm components of the outburst peaks 
of our PPS.
Particularly, an almost constant temperature 
and a shrinking BB radius during the outburst decay (Figs. \ref{pps2} 
and \ref{pps2b}). To our knowledge, the only other work that have 
performed a spectral analysis over long time-scales is the one of 
\citet{Albano2010}, who, in the framework of the twisted magnetospheric 
model (Thompson et al. 2002), used three modified BB to model the 
spectra, similar to the approach we based our work on (see text for 
details). Taking into account that \citet{Albano2010} did not 
perform a spectral fit, but obtained the physical parameters from 
synthetic pulse profiles, and, more importantly, that their values 
correspond to phase-average spectra, it is difficult to make a direct 
comparison with our results. Nonetheless, the thermal evolution of the 
BB modeled on \citet{Albano2010} is very similar to the one we infer: 
a constant warm component and a slightly decreasing value of the hot 
component temperature, while still consistent with a constant within 
the $1\sigma$ errors. Likewise, the radius evolution of the hot 
component in \citet{Albano2010} is as well very similar to the one 
we infer for the peaks: it significantly decreases throughout the 
outburst decay, ultimately disappearing in about 700 d.
On the other hand, in \citet{Albano2010} the warm component 
increases in size throughout the outburst, in contrast with what we 
infer in this work. Yet, the analysis of another magnetar 
considered by \citet{Albano2010}, XTE J1810-197, show the same decay 
trend we see in the outburst peaks: the hot and warm components keep 
an almost constant temperature and fade away in size, leaving the 
star emitting at the quiescence temperature towards the end of the 
outburst decay (in the case of a third, constant, ``cool'' BB 
temperature, see Albano et al. 2010 for details). 

Prior to the outburst, the pulse profile of the \src\ was single-peaked. 
The outburst strongly changes the observed pulse profile, and a 
three-peaked structure is clearly seen from the onset and during most 
of the outburst. Nevertheless, as the outburst decays, the pulse 
profile evolves and towards the end of the 2006 campaign, as the 
luminosity decreases, and \src\ returns to its quiescence level, the 
pulse profile ``returns'' to a single-peaked structure. The remaining 
peak correspond to the peak 1, and it is plausibly to assume that it 
correspond to the quiescence single peak. 



\begin{figure}
  \centering
  \includegraphics[width=0.35\textwidth, angle=270]{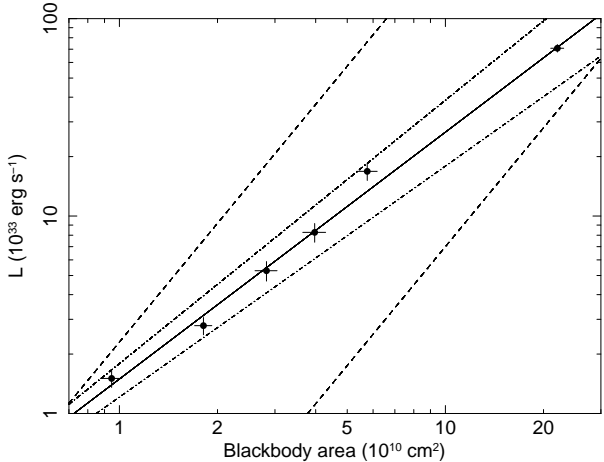}
  \caption{Luminosity versus emitting area of the warm BB
    component of the third peak (See Fig. \ref{pps1} and \ref{timing} for reference).
    The dashed lines represent the $L \propto A^{2}$ of simple 
    untwisting magnetosphere models, see Beloborodov, (2009).
    The solid line is a PL fit to the data which yields 
    $L \propto A^{1.25}$. The dot-dashed lines represent the 
    $3\sigma$ uncertainty of the fit (and correspond to
    $L \propto A^{1.17}$ and $L \propto A^{1.34}$).}
  \label{L.vs.A}
\end{figure}

The radius shrinking decay picture fits into the untwisting 
magnetosphere (UM) model (Beloborodov, 2009), where current-carrying 
``j-bundles'' with twisted magnetic fields gradually shrink. 
A simple UM model predicts the relation $L \propto A^2$ between 
the luminosity and the emitting area (see Beloborodov 2009, 
Equation 48); in Fig. \ref{L.vs.A} we compare the emitting area 
evolution with luminosity decay for the warm component of the
third peak with this modeled relationship. The PL fits
well the data but our analysis suggests a somehow flatter 
dependence then expected by the simple model, see Fig. \ref{L.vs.A}. 

An important issue is that this interpretation is model-dependent
and modeling the data with other spectral components can potentially 
yield a different picture. 
Indeed, other non-purely thermal models may also fit well the data,  
for instance, a BB+PL model and a resonant cyclotron Scattering 
(RCS) model (Rea et al. 2008) also fit the data acceptably.
For instance, the fit for 060922, the best observation in terms of 
signal-to-noise ratio, has $\chi^{2}_{red} = 0.97252$ (130 dof) and 
$\chi^{2}_{red} = 1.0993$ (130 dof), for the BB+PL and the RCS models
respectively. While our 1+2BB model has $\chi^{2}_{red} = 1.0210$ (129 dof).

On the other hand, independently from the spectral analysis the non-zero, 
negative second period derivative can also be accounted for within the 
UM model, as the magnetic field untwists, the spin-down torques diminish, 
effectively lowering the spin-down rate. However, there may be other 
explanations to the observed second period derivative, as wind braking, 
see e.g. \citet{tong2013}.

\begin{figure}
\includegraphics[width=6.1cm,angle=270]{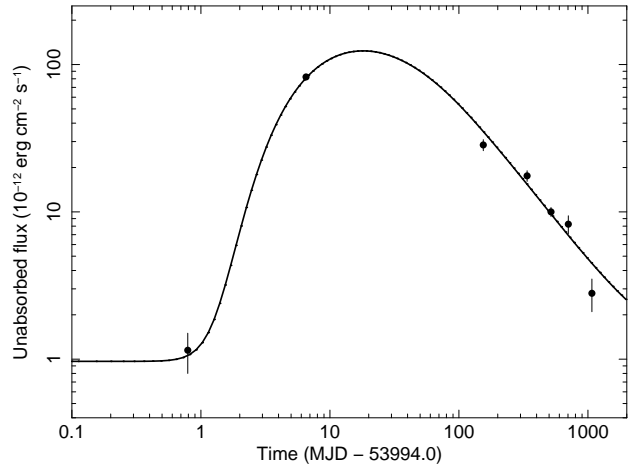}
\caption{Time evolution of the 0.5-10 keV unabsorbed 
flux, compared to the predicted light curve of the model
discussed in the text. 
\label{outburstmodel}}
\end{figure}

Furthermore, we also compared the observations of the 2006 outburst 
with the theoretical model presented in Pons \& Rea (2012). 
The pre-outburst model is taken to be the evolved NS that fits the 
present observational constraints. Then we assume that the source 
undergoes a sudden starquake, possibly with internal magnetic 
re-connection, which we model by the injection of energy 
($\approx 10^{25}$-$10^{26}$ erg cm$^{-3}$) in the thin layer between 
two variable densities. We found a good agreement with the luminosity 
data when the energy is deposited between $\rho = 2 \times 10^{9}$
and $2 \times 10^{10}$ g cm$^{-3}$, precisely in the transition region between the outer crust
and the liquid envelope, which may be a hint that the energy is provided not only by elastic energy
stored in the solid crust but also by magnetic re-connection in the liquid layer.
The time evolution of the unabsorbed flux in the 0.5-10 keV band
for this particular model is shown in Fig. \ref{outburstmodel} and superimposed to the measured flux values.
The total injected energy was of $2 \times 10^{43}$ erg. We note that the last observation seems
to show a smaller flux than the prediction of the theoretical model. Interestingly, the same effect
has been observed and discussed for \sgr\ (SGR 0418), where the sudden decrease of the flux after
300 days is not well understood (Rea et al. 2013). The occurrence of a second outburst soon after
this last data point, does not allow to determine if the source had fully recovered its quiescence state
or it was still cooling down.

Note that a short-term ($\sim 10$ d) rise in temperature early in 
the outburst onset, reported by An et al. (2013), which may be expected 
from crustal cooling models is out of the long-term evolution scope of this paper.


\subsection{Magnetorotational Evolution}
\label{magnetorot}

\begin{figure}
\includegraphics[width=8.3cm,angle=0]{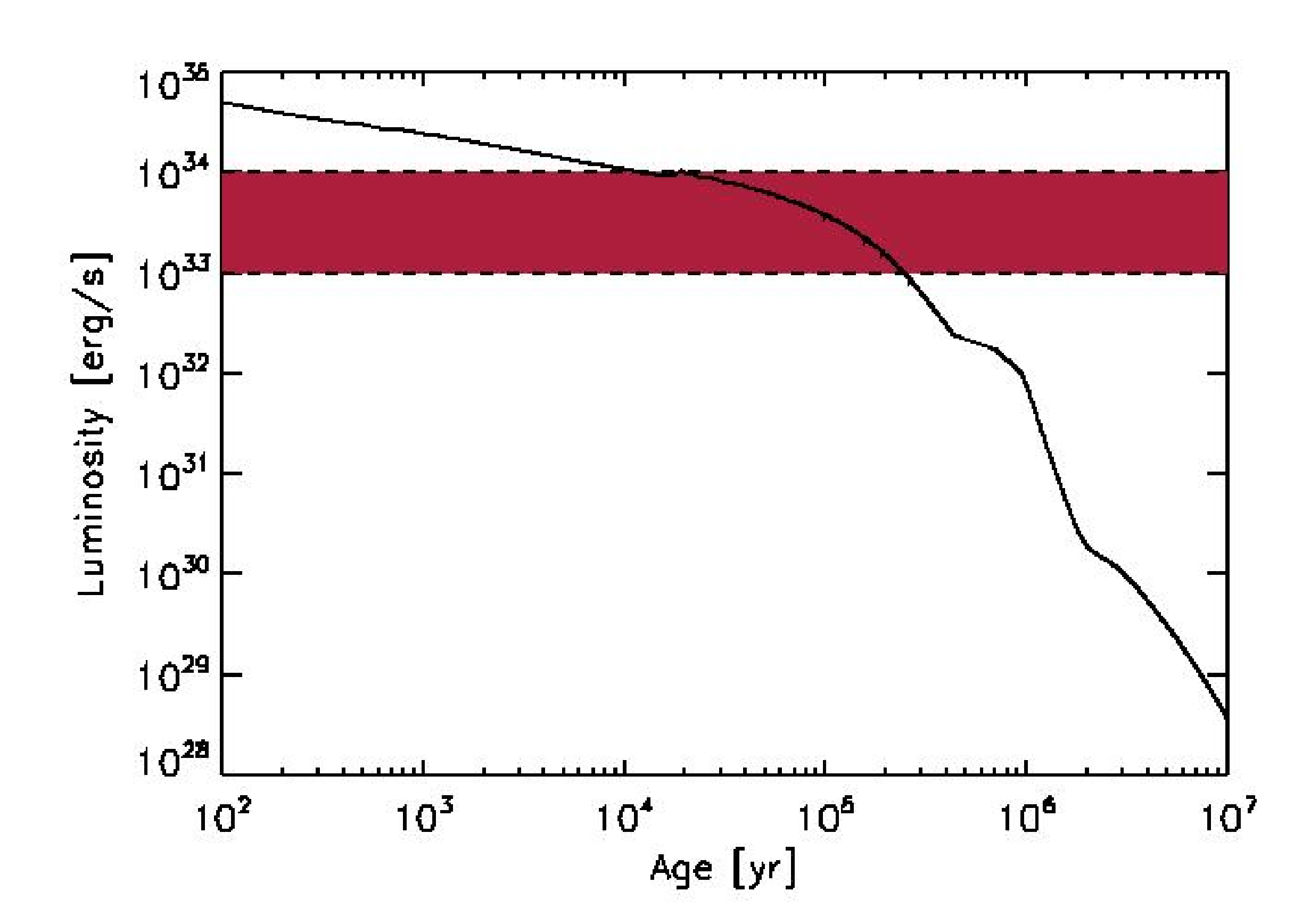}
\includegraphics[width=8.3cm,angle=0]{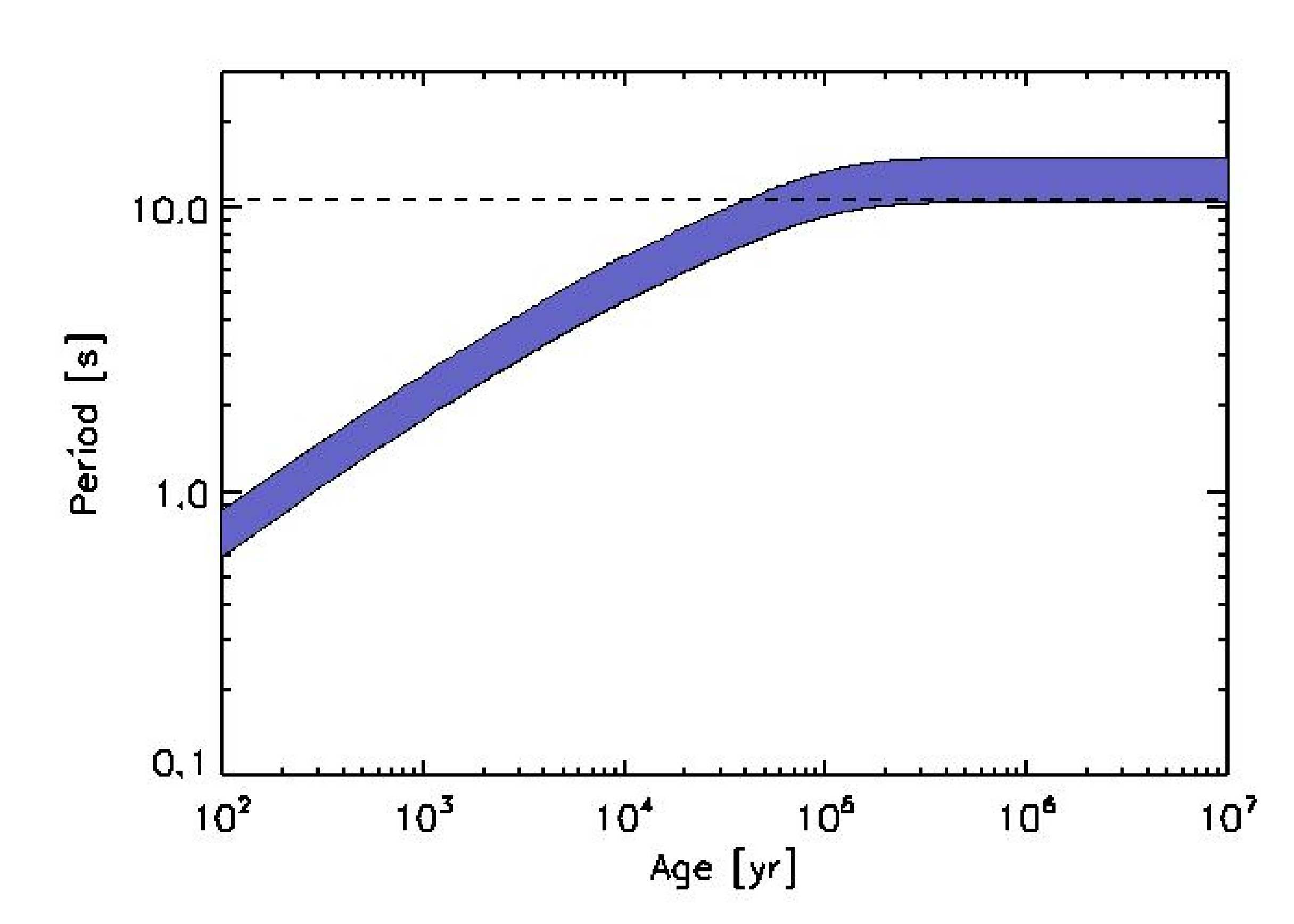}
\includegraphics[width=8.3cm,angle=0]{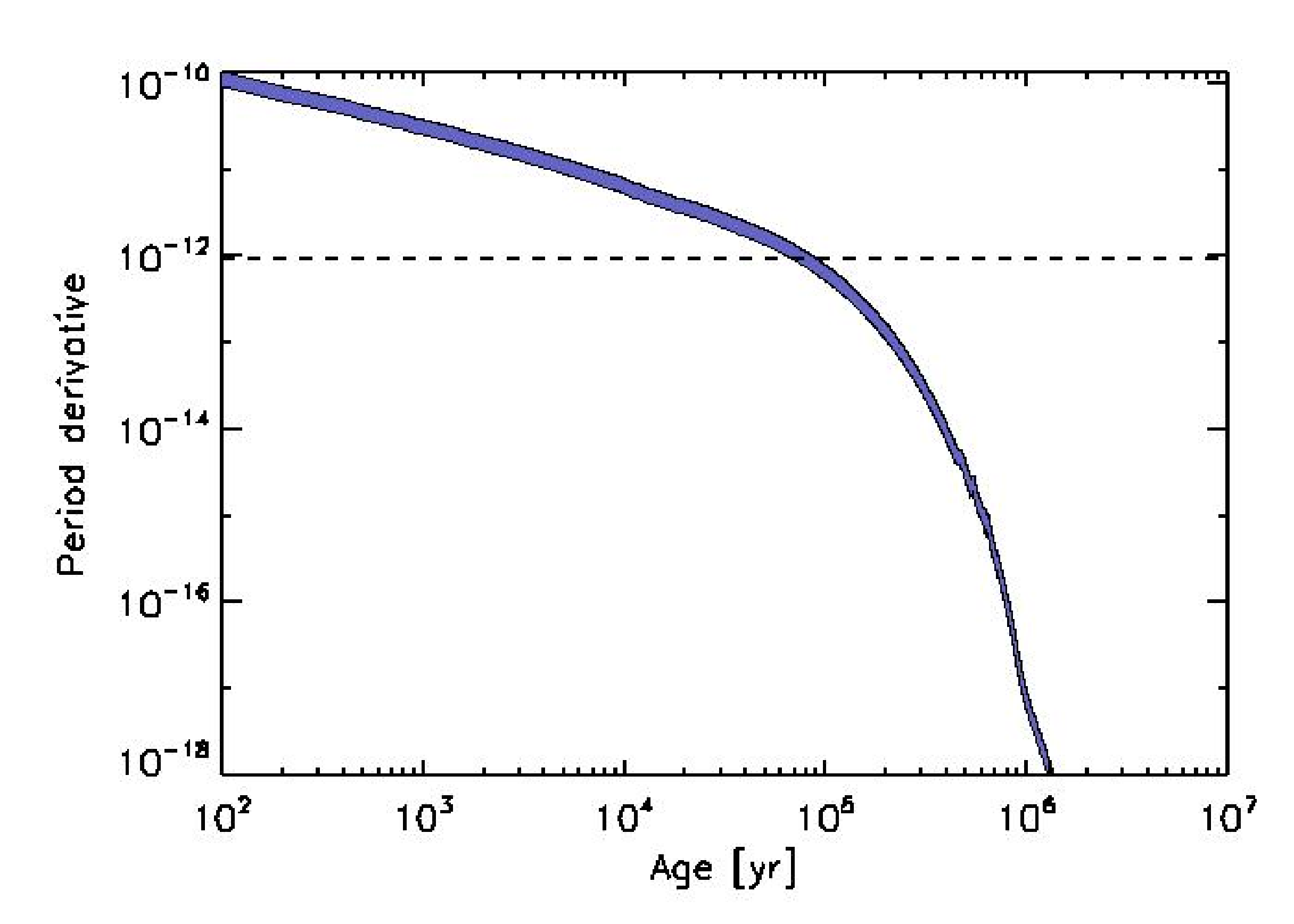}
\caption{From top  to bottom, the evolution of the
luminosity, period and period derivative
according to the model discussed in the text, compared to the
measured values.
\label{mag-rot-evol}}
\end{figure}

As previously done for other magnetars (SGR 0418 and Swift J1822.3-1606; 
see Turolla et al. 2011 and Rea et al. 2012b, 2013) we explore if the 
magnetothermal evolution of a NS born with standard magnetar conditions 
can lead to objects with properties compatible with those of CXOU J1647 
at the present age. We performed some runs using state-of-the-art 
magnetothermal evolution codes (see Vigan\`o et al. 2013) assuming a $1.4M_\odot$ 
NS with radius $R=11.6$ km, a short initial period of 10 ms 
and initial, purely dipolar field of $B=1.5\times 10^{14}$ G. 
In the resulting scenario, the Hall term reorganize the internal field, producing
a toroidal component of the same strength as the poloidal 
one on a relatively short time-scale (within a few kyr).
We show in Fig. \ref{mag-rot-evol} the evolution of the 
luminosity, period $P$ and period derivative $\dot P$. 
The latter two quantities are obtained, from the value of the magnetic field
at the equator $B(t)$, by numerical  integration of the expression (\citealt{Spitkovsky2006})
\begin{equation}
\label{spindown}
P \dot{P} \simeq \frac{4B_e^2R^6 \pi^2}{I c^3} (1+\sin^2{\alpha})
\end{equation}
where $I$ is the effective moment of inertia of the star, $\alpha$ is 
the angle between the rotational and the magnetic axis and $c$ is the speed of light. 
The shaded blue area in the figure includes the uncertainty in the angle $\alpha$.
Indeed the properties of CXOU J1647 are recovered by this model at an age between 65 and 90 kyr, 
about half the spin-down age, which suggests that the magnetic field has not experienced 
dramatic changes over time.

Although the components of the internal initial field $B_{tor}(t=0)$ can be
varied to some extent, this would not change our results unless the toroidal field contains
most of the magnetic energy ($> 90 \%$), as discussed in Vigan\`o et al. (2013). Moderate
values of the initial toroidal field (or higher order poloidal multipoles), are unconstrained
and will result in very similar properties at the present age.
We can also estimate the current outburst rate of this source following the procedure of 
Perna \& Pons (2011), which gives $\lesssim 10^{-2}$ yr$^{-1}$. Therefore, within our model, 
the occurrence of a second outburst in 2009, three years after the first outburst, 
must be connected to the first event. Since the second outburst is less powerful, 
the pulse profile after it closely resembles the pulse profile after the 
initial (2006) one, and the pulsed fraction does not present a strong change 
(as the sharp fall after the 2006 outburst onset), but rather seems to follow the rising trend
seen during the outburst (see Fig. \ref{PF}); it may be speculated 
that there is a connection between them, of the kind ’main event + 
sequel’, which could reconcile the model with the observations.

\subsection*{Acknowledgments}
The authors would like to thank Tolga Guver for his excellent suggestions during the 
reviewing process that greatly improved the manuscript.
DV was supported by the grants AYA 2010-21097-C03-02, ACOMP/2012/135, AYA 2012-39303 and
SGR 2009-811.
NR was supported by a Vidi NWO grant, Ramon y Cajal Fellowship, AYA 2012-39303 and SGR 2009-811.

\nocite{*} 
\bibliography{bib}

\begin{thebibliography}{}

\bibitem[\protect\citeauthoryear{{From outburst to quiescence: the decay of the
  transient AXP XTE J1810-197}}{Ng2}{}]{Ng2011}


\bibitem[\protect\citeauthoryear{{Exciting the magnetosphere of the magnetar
  CXOU J164710.2-455216 in Westerlund 1}}{Kou}{}]{Kouveliotou2003}


\bibitem[\protect\citeauthoryear{Albano, Turolla, Israel, Zane, Nobili \&
  Stella}{Albano et~al.}{2010}]{Albano2010}
Albano A.,  Turolla R.,  Israel G.~L.,  Zane S.,  Nobili L.,    Stella L.,
  2010, ApJ, 722, 788

\bibitem[\protect\citeauthoryear{{Alpar}, {{\c C}al{\i}{\c s}kan} \&
  {Ertan}}{{Alpar} et~al.}{2013}]{Alpar2013}
{Alpar} M.~A.,  {{\c C}al{\i}{\c s}kan} {\c S}.,    {Ertan} {\"U}.,  2013, in
  {Zhang} C.~M.,  {Belloni} T.,  {M{\'e}ndez} M.,   {Zhang} S.~N.,  eds, IAU
  Symposium Vol.~290 of IAU Symposium, {Fallback Disks, Magnetars and Other
  Neutron Stars}.
pp 93--100

\bibitem[\protect\citeauthoryear{An, Kaspi, Archibald \& Cumming}{An
  et~al.}{2013}]{An2013}
An H.,  Kaspi V.~M.,  Archibald R.,    Cumming A.,  2013, ApJ, 763, 82

\bibitem[\protect\citeauthoryear{{Baumgartner}, {Burrows}, {Cummings},
  {Gehrels}, {Gronwall}, {Holland}, {Kennea}, {Mangano}, {Markwardt},
  {Marshall}, {Racusin}, {Sbarufatti}, {Sonbas} \& {Ukwatta}}{{Baumgartner}
  et~al.}{2011}]{Baumgartner2011}
{Baumgartner} W.~H.,  {Burrows} D.~N.,  {Cummings} J.~R.,  {Gehrels} N.,
  {Gronwall} C.,  {Holland} S.~T.,  {Kennea} J.~A.,  {Mangano} V.,  {Markwardt}
  C.~B.,  {Marshall} F.~E.,  {Racusin} J.~L.,  {Sbarufatti} B.,  {Sonbas} E.,
   {Ukwatta} T.~N.,  2011, GRB Coordinates Network, 12359, 1

\bibitem[\protect\citeauthoryear{Beloborodov}{Beloborodov}{2009}]{Beloborodov2009}
Beloborodov A.~M.,  2009, ApJ, 703, 1044

\bibitem[\protect\citeauthoryear{Bernardini \& et al.}{Bernardini \&
  et~al.}{2009}]{Bernardini2009}
Bernardini F.,  et al. 2009, A\&A, 498, 195

\bibitem[\protect\citeauthoryear{Camilo, Ransom, Halpern \& Reynolds}{Camilo
  et~al.}{2007}]{Camilo2007}
Camilo F.,  Ransom S.~M.,  Halpern J.~P.,    Reynolds J.,  2007, ApJ, 666, L93

\bibitem[\protect\citeauthoryear{Camilo, Ransom, Halpern, Reynolds, Helfand,
  Zimmerman \& Sarkissian}{Camilo et~al.}{2006}]{Camilo2006}
Camilo F.,  Ransom S.~M.,  Halpern J.~P.,  Reynolds J.,  Helfand D.~J.,
  Zimmerman N.,    Sarkissian J.,  2006, Nature, 442, 892

\bibitem[\protect\citeauthoryear{Campana \& Israel}{Campana \&
  Israel}{2006}]{Campana2006}
Campana S.,  Israel G.~L.,  2006, The Astronomer's Telegram, 893

\bibitem[\protect\citeauthoryear{Crowther, Hadfield, Clark, Negueruela \&
  Vacca}{Crowther et~al.}{2006}]{Crowther2006a}
Crowther P.~a.,  Hadfield L.~J.,  Clark J.~S.,  Negueruela I.,    Vacca W.~D.,
  2006, MNRAS, 372, 1407

\bibitem[\protect\citeauthoryear{{Dall'Osso}, {Israel}, {Stella}, {Possenti} \&
  {Perozzi}}{{Dall'Osso} et~al.}{2003}]{DallOsso2003}
{Dall'Osso} S.,  {Israel} G.~L.,  {Stella} L.,  {Possenti} A.,    {Perozzi} E.,
   2003, ApJ, 599, 485

\bibitem[\protect\citeauthoryear{Duncan \& Thompson}{Duncan \&
  Thompson}{1992}]{Duncan1992}
Duncan R.~C.,  Thompson C.,  1992, ApJ, 392, L9

\bibitem[\protect\citeauthoryear{Horvath}{Horvath}{2005}]{Horvath2005}
Horvath J.~E.,  2005, Modern Physics Letters A, 20, 2799

\bibitem[\protect\citeauthoryear{Ibrahim \& et al.}{Ibrahim \&
  et~al.}{2004}]{Ibrahim2004}
Ibrahim A.~I.,  et al. 2004, ApJ, 609, L21

\bibitem[\protect\citeauthoryear{Israel, Campana, Dall’Osso, Muno, Cummings,
  Perna \& Stella}{Israel et~al.}{2007}]{Israel2007}
Israel G.~L.,  Campana S.,  Dall’Osso S.,  Muno M.~P.,  Cummings J.,  Perna
  R.,    Stella L.,  2007, ApJ, 664, 448

\bibitem[\protect\citeauthoryear{Israel, Esposito \& Rea}{Israel
  et~al.}{2011}]{Israel2006}
Israel G.~L.,  Esposito P.,    Rea N.,  2011, The Astronomer's Telegram, 3653

\bibitem[\protect\citeauthoryear{Israel \& et al.}{Israel \&
  et~al.}{2004}]{Israel2004}
Israel G.~L.,  et al. 2004, ApJ, 603, L97

\bibitem[\protect\citeauthoryear{Israel \& et al.}{Israel \&
  et~al.}{2009}]{Israel2009}
Israel G.~L.,  et al. 2009, The Astronomer's Telegram, 1909

\bibitem[\protect\citeauthoryear{{Malheiro}, {Rueda} \& {Ruffini}}{{Malheiro}
  et~al.}{2012}]{Malheiro2012}
{Malheiro} M.,  {Rueda} J.~A.,    {Ruffini} R.,  2012, PASJ, 64, 56

\bibitem[\protect\citeauthoryear{Mereghetti}{Mereghetti}{2008}]{Mereghetti2008}
Mereghetti S.,  2008, A\&ARv, 15, 225

\bibitem[\protect\citeauthoryear{Mereghetti}{Mereghetti}{2013}]{Mereghetti2013}
Mereghetti S.,  2013, Brazilian Journal of Physics

\bibitem[\protect\citeauthoryear{Muno \& et al.}{Muno \&
  et~al.}{2006}]{Muno2006}
Muno M.~P.,  et al. 2006, ApJ, 636, L41

\bibitem[\protect\citeauthoryear{Muno, Gaensler, Clark, de Grijs, Pooley,
  Stevens \& Zwart}{Muno et~al.}{2007}]{Muno2007}
Muno M.~P.,  Gaensler B.~M.,  Clark J.~S.,  de Grijs R.,  Pooley D.,  Stevens
  I.~R.,    Zwart S. F.~P.,  2007, MNRAS: Letters, 378, L44

\bibitem[\protect\citeauthoryear{{Negueruela}, {Clark} \&
  {Ritchie}}{{Negueruela} et~al.}{2010}]{Negueruela2010}
{Negueruela} I.,  {Clark} J.~S.,    {Ritchie} B.~W.,  2010, AAp, 516, A78

\bibitem[\protect\citeauthoryear{Nobili, Turolla \& Zane}{Nobili
  et~al.}{2008}]{ntz2008}
Nobili L.,  Turolla R.,    Zane S.,  2008, MNRAS, 386, 1527

\bibitem[\protect\citeauthoryear{{Ouyed}, {Leahy} \& {Niebergal}}{{Ouyed}
  et~al.}{2010}]{Ouyed2010}
{Ouyed} R.,  {Leahy} D.,    {Niebergal} B.,  2010, AAp, 516, A88

\bibitem[\protect\citeauthoryear{{Paczynski}}{{Paczynski}}{1990}]{Paczynski1990}
{Paczynski} B.,  1990, ApJL, 365, L9

\bibitem[\protect\citeauthoryear{Perna \& Pons}{Perna \&
  Pons}{2011}]{Perna2011}
Perna R.,  Pons J.~A.,  2011, ApJ, 727, L51

\bibitem[\protect\citeauthoryear{Pons \& Rea}{Pons \& Rea}{2012}]{Pons2012}
Pons J.~A.,  Rea N.,  2012, ApJ, 750, L6

\bibitem[\protect\citeauthoryear{Rea \& Esposito}{Rea \&
  Esposito}{2011}]{Rea2011}
Rea N.,  Esposito P.,  2011, High-Energy Emission from Pulsars and their
  Systems, Astrophysics and Space Science Proceedings, p.~247

\bibitem[\protect\citeauthoryear{Rea \& et al.}{Rea \& et~al.}{2012}]{Rea2012a}
Rea N.,  et al. 2012, ApJ, 754, 27

\bibitem[\protect\citeauthoryear{Rea \& et al.}{Rea \& et~al.}{2013}]{Rea2013}
Rea N.,  et al. 2013, ApJ, 770, 65

\bibitem[\protect\citeauthoryear{Rea, Pons, Torres \& Turolla}{Rea
  et~al.}{2012}]{Rea2012}
Rea N.,  Pons J.,  Torres D.~F.,    Turolla R.,  2012, ApJ, 748, L12

\bibitem[\protect\citeauthoryear{Rea, Zane, Turolla, Lyutikov \& G\"{o}tz}{Rea
  et~al.}{2008}]{Rea2008}
Rea N.,  Zane S.,  Turolla R.,  Lyutikov M.,    G\"{o}tz D.,  2008, ApJ, 686,
  1245

\bibitem[\protect\citeauthoryear{{Spitkovsky}}{{Spitkovsky}}{2006}]{Spitkovsky2006}
{Spitkovsky} A.,  2006, ApJL, 648, L51

\bibitem[\protect\citeauthoryear{Thompson \& Duncan}{Thompson \&
  Duncan}{1995}]{ThompsonChristopher1995}
Thompson C.,  Duncan R.~C.,  1995, MNRAS, 275, 255

\bibitem[\protect\citeauthoryear{Thompson, Lyutikov \& Kulkarni}{Thompson
  et~al.}{2002}]{Thompson2002}
Thompson C.,  Lyutikov M.,    Kulkarni S.~R.,  2002, ApJ, 574, 332

\bibitem[\protect\citeauthoryear{Tiengo \& et al.}{Tiengo \&
  et~al.}{2013}]{Tiengo2013}
Tiengo A.,  et al. 2013, Nature, 500, 312

\bibitem[\protect\citeauthoryear{{Tong}, {Xu}, {Song} \& {Qiao}}{{Tong}
  et~al.}{2013}]{tong2013}
{Tong} H.,  {Xu} R.~X.,  {Song} L.~M.,    {Qiao} G.~J.,  2013, ApJ, 768, 144

\bibitem[\protect\citeauthoryear{Turolla \& P.}{Turolla \&
  P.}{2013}]{turolla2013}
Turolla R.,  P. E.,  2013, preprint (arXiv:1303.6052)

\bibitem[\protect\citeauthoryear{Turolla, Zane, Pons, Esposito \& Rea}{Turolla
  et~al.}{2011}]{Turolla2011a}
Turolla R.,  Zane S.,  Pons J.~a.,  Esposito P.,    Rea N.,  2011, ApJ, 740,
  105

\bibitem[\protect\citeauthoryear{{van Paradijs, J.; Taam, R. E.; van den
  Heuvel}}{{van Paradijs, J.; Taam, R. E.; van den
  Heuvel}}{1995}]{Paradijs1995}
{van Paradijs, J.; Taam, R. E.; van den Heuvel} E. P.~J.,  1995, A\&A, 299, L41

\bibitem[\protect\citeauthoryear{Vigan\`o, Rea, Pons, Perna, Aguilera \&
  Miralles}{Vigan\`o et~al.}{2013}]{Vigano2013}
Vigan\`o D.,  Rea N.,  Pons J.~A.,  Perna R.,  Aguilera D.~N.,    Miralles
  J.~A.,  2013, MNRAS, 434, 123

\bibitem[\protect\citeauthoryear{Woods \& et al.}{Woods \&
  et~al.}{2000}]{Woods2000}
Woods P.~M.,  et al. 2000, ApJ, 535, L55

\bibitem[\protect\citeauthoryear{Woods, Kaspi, Gavriil \& Airhart}{Woods
  et~al.}{2011}]{Woods2011}
Woods P.~M.,  Kaspi V.~M.,  Gavriil F.~P.,    Airhart C.,  2011, ApJ, 726, 37

\bibitem[\protect\citeauthoryear{Xu}{Xu}{2007}]{Xu2007}
Xu R.,  2007, Advances in Space Research, 40, 1453

\end{thebibliography}
\bibliographystyle{mn2e}

\end{document}